\documentclass[12pt]{article}
\usepackage{epsfig}
\usepackage{amsmath}
\usepackage[section]{placeins}
\makeatletter
\usepackage{authblk}

\def\@author{}
\renewcommand\@author{\ifx\AB@affillist\AB@empty\AB@author\else
	\ifnum\value{affil}>\value{Maxaffil}\def\rlap##1{##1}%
    \AB@authlist \;on behalf of the CMS Collaboration\\[\affilsep]\AB@affillist
    \else  \AB@authors\fi\fi}
\makeatother

\title{Search for new resonances decaying into boosted W, Z and H bosons at CMS}
\author[1]{Michael Krohn}
\author[2]{Caterina Vernieri}
\affil[1]{University of Colorado Boulder, USA}
\affil[2]{Fermilab, IL, USA}
\date{}

\begin{document}

\maketitle






\begin{abstract}
An overview of the searches for new heavy resonances decaying to standard model (SM) bosons at the TeV mass scale is presented. Results are based on data corresponding to an integrated luminosity up to about $36\; \mathrm{fb}^{-1}$ recorded in proton-proton collisions at $\sqrt{s}=13$ TeV with the CMS detector at the Large Hadron Collider at CERN. The bosons coming from the resonance decay can be W, Z, or the SM Higgs. For very heavy resonances, bosons are produced with momentum considerably higher than their mass, modifying in a very appreciable way the event topology. The quarks originated from the hadronic decay of the SM bosons will be collimated into a smaller area such that they are clustered within a single large cone jet. Dedicated reconstruction techniques are used to distinguish the merged decay products of W, Z and H bosons produced with high transverse momentum, from jets that originate from single partons.
\end{abstract}

\section{Introduction}

The "hierarchy problem", meaning the several orders of magnitude discrepancy between the gravitational and the weak forces, is an open question of the standard model (SM) and many theories beyond the standard model (BSM) aim to explain it. Possible BSM scenarios predict the existence of heavy, narrow width particles that can couple to a pair of standard model bosons, specifically the W and Z vector bosons (V) or the Higgs boson (H). For example, models with a warped extra dimension (WED), as proposed by Randall and Sundrum~\cite{WEDs}, predict the existence of new particles, such as the spin-0 radion and the spin-2 first Kaluza-Klien excitation of the graviton. There is also the heavy vector triplet (HVT) model~\cite{HVT} which extends the SM by introducing a triplet of heavy vector bosons. 

The decay of new, heavy particles (X) to pairs of V or H bosons is searched for in the all hadronic final state in multiple searches at CMS. The hadronic final state has the largest branching fraction of V or H decays and thus these searches are the most sensitive at high resonance mass. The topology of the final state is constrained by $M_{X}/(M_{V_{1}(H_{1})}+M_{V_{2}(H_{2})}) >> 1$, where $M_{X}$ is the mass of the resonant particle, and defines the so-called boosted regime, in which each V or H is produced with a large momentum and its decay products are collimated along its direction of motion. Each boson is thus reconstructed as a large cone size jet (fat jet) with a mass compatible with the V or H mass that is searched for. The boson candidates are selected by employing jet substructure and b tagging techniques to exploit both the composite nature and the flavor of the jets.

This document presents the search for heavy, narrow resonances decaying to VV, VH, and HH pairs in the all hadronic final state. These analyses use proton-proton collision data at $\sqrt{s}=13$ TeV collected by the CMS detector in 2016~\cite{CMS}, corresponding to an integrated luminosity of $35.9\;\mathrm{fb}^{-1}$. 

\section{Boosted Boson Identification}\label{BoostedBosonIdentification}

The angular separation of the two quarks that decay from a SM boson is proportional to $2M_{V(H)}/p_{T}$, where $p_{T}$ is the transverse momentum of the V(H). When the $p_{T}$ of the boson is greater than $\sim250$ GeV it is more efficient to reconstruct the boson as a fat jet rather than two small-cone jets. At CMS these large-cone jets have a $\Delta R = 0.8$ and are reconstructed using the anti-$k_{T}$ algorithm~\cite{Antikt} (AK8) and thus these searches for pairs of SM bosons look for two AK8 jets. To distinguish a SM boson jet from a jet that originates from a quark or gluon jet observables are used. In order to mitigate the effect of pileup on jet observables, the pileup per particle identification (PUPPI)~\cite{PUPPI} algorithm is applied. This method uses local shape information, event pileup properties, and tracking information together in order to compute a weight describing the degree to which a particle is pileup-like. 

\begin{figure}[!htb]
\begin{center}
\includegraphics[height=2.5in]{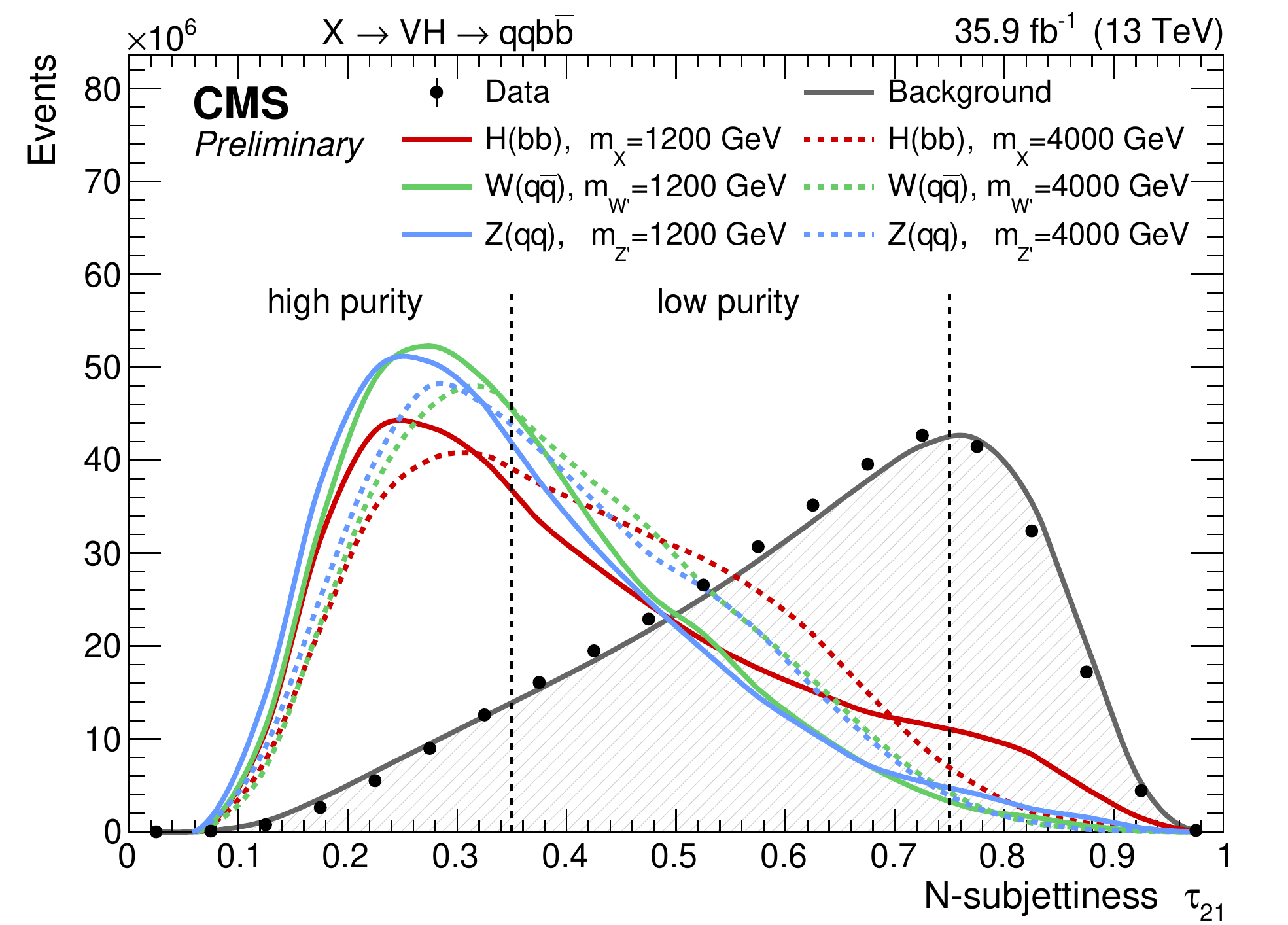}
\caption{Distribution of the N-subjettiness $\tau_{21}$ after the kinematic selections on the two jets, for data, simulated background, and signal. The distributions are normalized to the number of events observed in data.~\cite{VHSearch}.}
\label{fig:Tau21}
\end{center}
\end{figure}

The "N-subjetttiness" algorithm~\cite{Nsubjettiness} is used to determine the consistency of the jet with two substructures from the two-pronged V(H) $\rightarrow$ $\mathrm{q\bar{q}}$($\mathrm{b\bar{b}}$) decay. The ratio of the N-subjettiness observables $\tau_{21} = \tau_{2}/\tau_{1}$ is found to be a good discriminator between jets with two prongs from jets with a single prong, as can be seen in Figure~\ref{fig:Tau21}\cite{VHSearch}.

\begin{figure}[!htb]
\begin{center}
\includegraphics[height=2in]{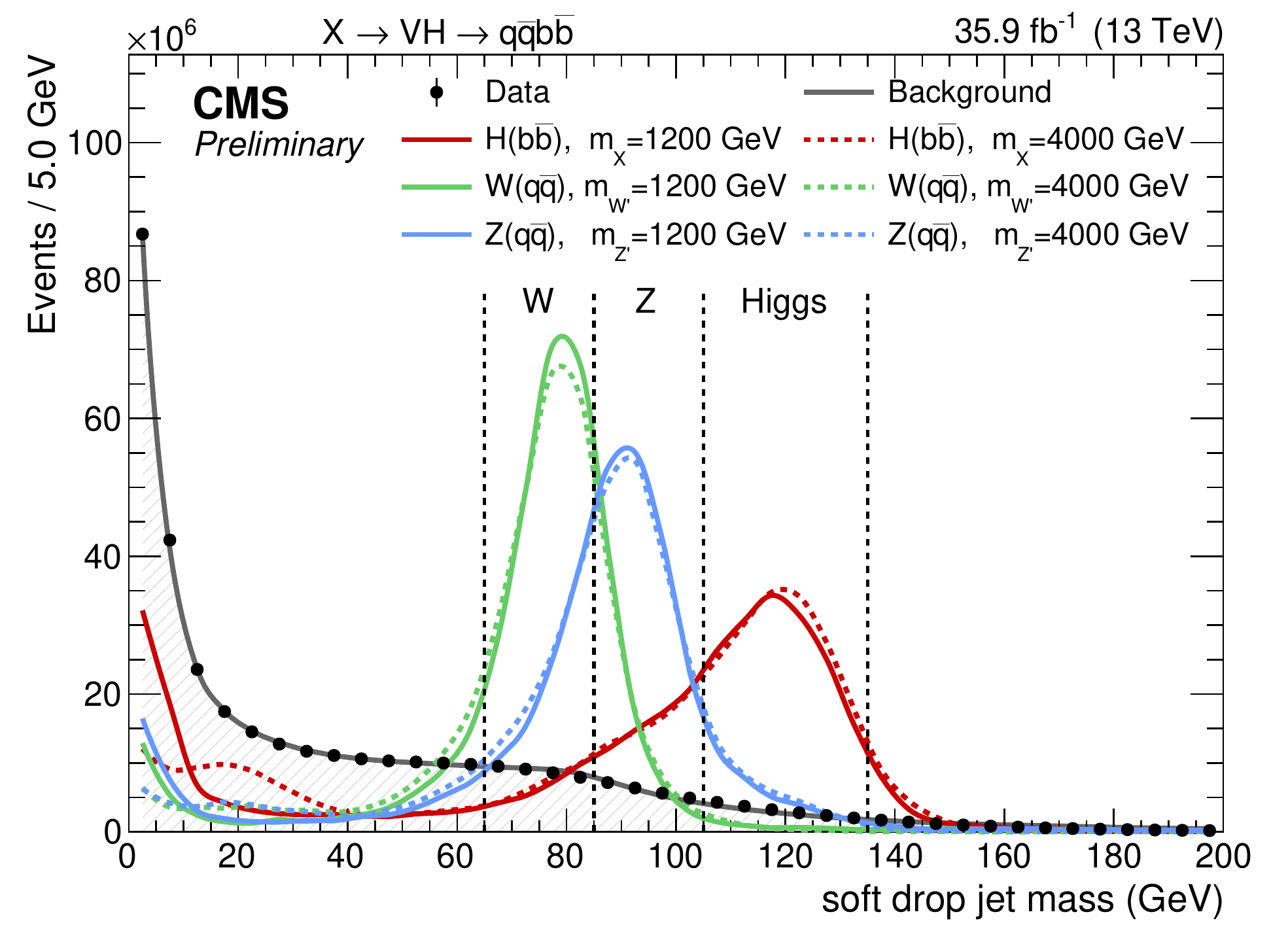}
\includegraphics[height=2in]{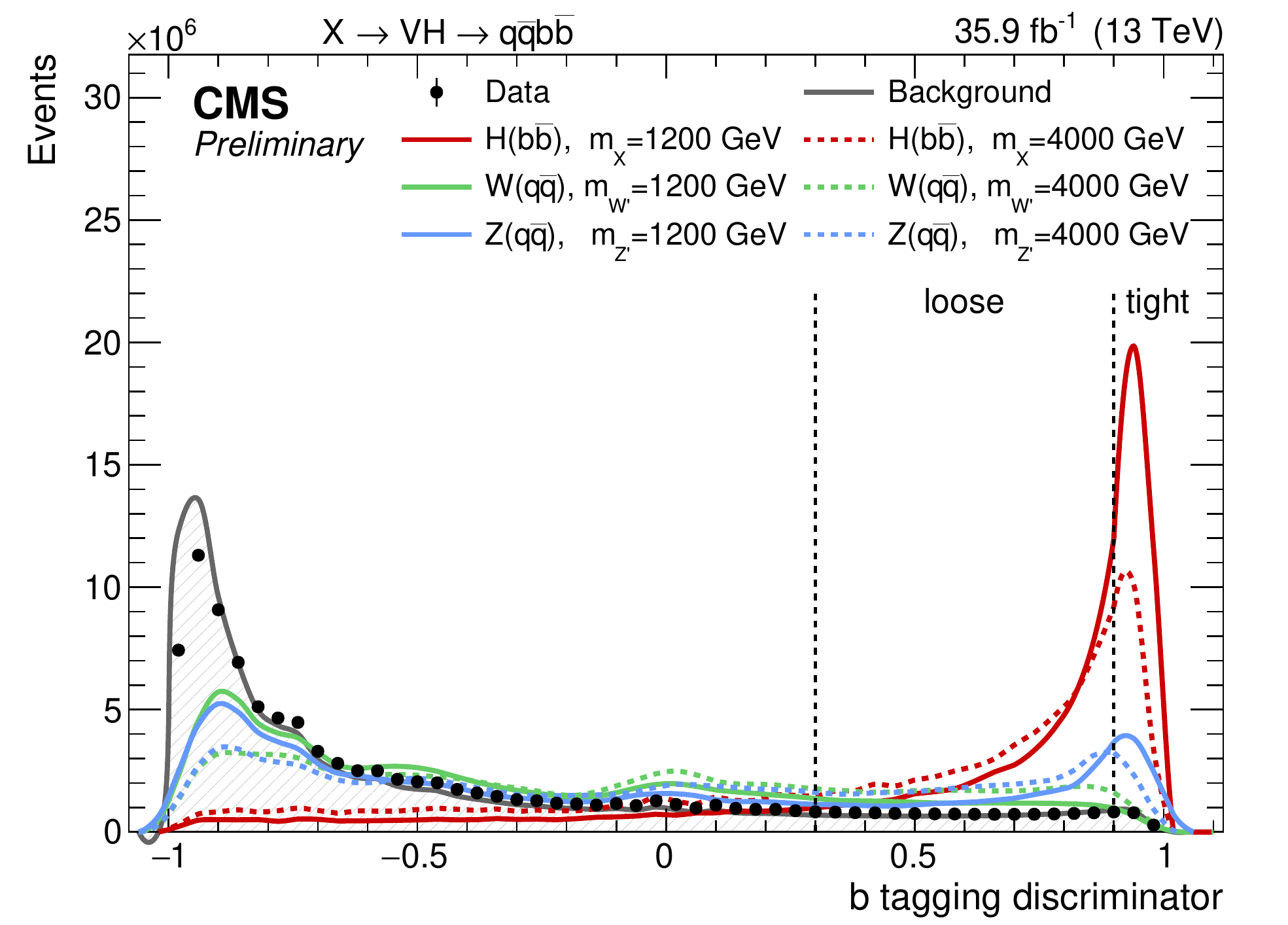}
\caption{
Distribution of the soft drop jet mass (left) and b tagging discriminator (right) after the kinematic selections on the two jets, for data, simulated background, and signal. The distributions are normalized to the number of events observed in data~\cite{VHSearch}.}
\label{fig:DoubleB}
\end{center}
\end{figure}

Grooming techniques are applied~\cite{JetGrooming} to remove soft and wide-angle radiation from the jet cone. Grooming tends to push the jet mass scale of the quark- and gluon-initiated (q/g) jets to lower values while preserving the hard scale of the boson jets. In CMS the soft drop~\cite{SDMass} algorithm is used with $\beta=1$, which corresponds to the modified mass drop procedure~\cite{ModifiedMass}. The soft drop mass peaks at the V or H boson mass for signal events and reduces the mass of background q/g jets. The distribution is shown in Figure~\ref{fig:DoubleB}~\cite{VHSearch} for simulated signal and background events.



In order to identify H or Z decaying to b quarks a double-b tagger discriminant~\cite{DoubleB} is employed. The double-b tagger exploits the presence of two hadronized b quarks inside the AK8 jet and their kinematics in relation to the jet substructure, namely the fact that the $B$ hadron flight directions are strongly correlated with the axes used to calculate the N-subjettiness observables. Several observables exploiting the $B$ hadron lifetime are used as input variables. The distribution of the double-b tagger for simulated signal and background events in Figure~\ref{fig:DoubleB}~\cite{VHSearch}.

\section{Search for X $\rightarrow$ VV($\mathrm{q\bar{q}}$$\mathrm{q\bar{q}}$)}

The search for a heavy resonant particle decaying to  a pair of V decaying hadronically is done by selecting events with two AK8 jets, each of them compatible with the boosted V tagging criteria described in Section~\ref{BoostedBosonIdentification}. The events are initially selected from events that pass online requirements based on the scalar sum of the transverse energy of all jets in the event ($H_{T}$) or the presence of one of more jets with loose substructure requirements. Online selections are fully efficient for resonance mass above 1 TeV. The two jets are also required to have an offline separation $\Delta\eta < 1.3$ because in background events the two jets are often well separated in $\eta$, especially at high resonance mass. In contrast, signal events that contain a heavy resonance decaying to two energetic  jets are characterized by a small separation of the two jets. The events are then separated into 6 orthogonal categories, 3 for the different mass compatibilities of the vector bosons (WW, WZ, ZZ) and 2 for the purity of the $\tau_{21}$ selection. The high purity (HP) category requires that both V candidates pass a high purity $\tau_{21}$ selection and the low purity (LP) category requires that one V candidate passes a high purity $\tau_{21}$ selection and the other candidate passes has a looser requirement. The LP category is used because it improves signal efficiency with an acceptable background contamination. 

The main variable used in the search for a VV resonance is the dijet invariant mass ($m_{jj}$). After all the event selection criteria are applied the dominant background is  QCD multijets. The background is directly estimated from data and the search for signal in the dijet mass spectrum is done using a "bump-hunt" approach. It assumes that the SM background is a smooth, monotonically decreasing distribution and any signal will appear as a narrow peak on top of the falling distribution. In each search category, a continuous function with either two or three parameters is used~\cite{VVSearch} to fit the background and the functions are chosen by performing a Fisher F-test~\cite{Fisher}. In the fits, based on a profile likelihood, the parameters and the normalization of the background in each category are free to float. Systematic uncertainties are treated as nuisance parameters and are profiled in the statistical interpretation~\cite{CLs,CLs2,CLs3,CLs4}. The fit functions are tested in QCD simulated events and data sidebands.

\begin{figure}[!htb]
\begin{center}
\includegraphics[height=3.in]{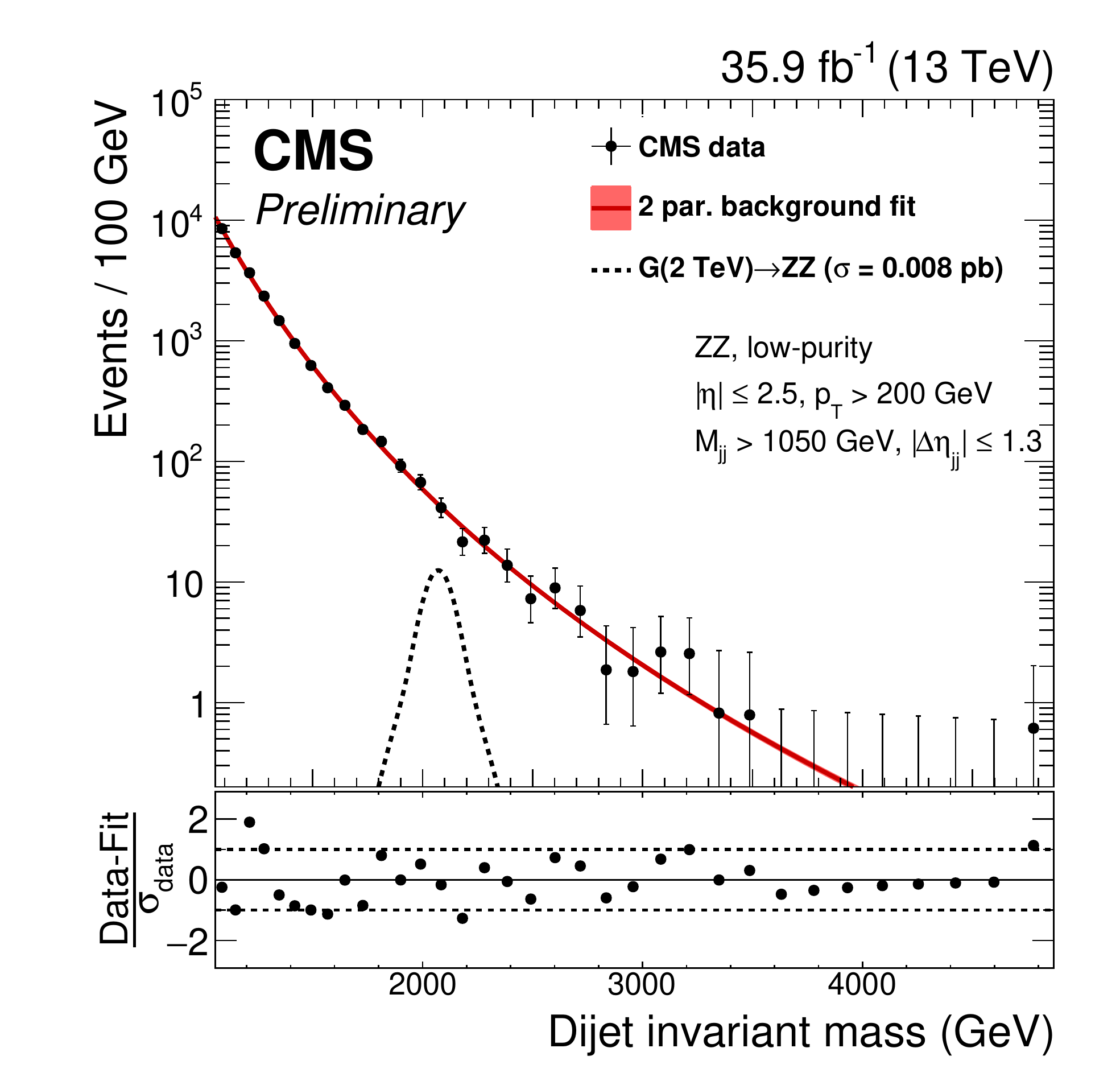}
\caption{
Dijet invariant distribution $m_{jj}$ of the two leading jets in the Z mass region for the low purity category. The preferred background-only fit is shown with an associated shaded band indicating the uncertainty. The differences between the data and the predicted background, divided by the data statistical uncertainty ($\sigma$) are shown in the lower panel~\cite{VVSearch}.}
\label{fig:VVBumpHunt}
\end{center}
\end{figure}
\begin{figure}[!htb]
\begin{center}
\includegraphics[height=2in]{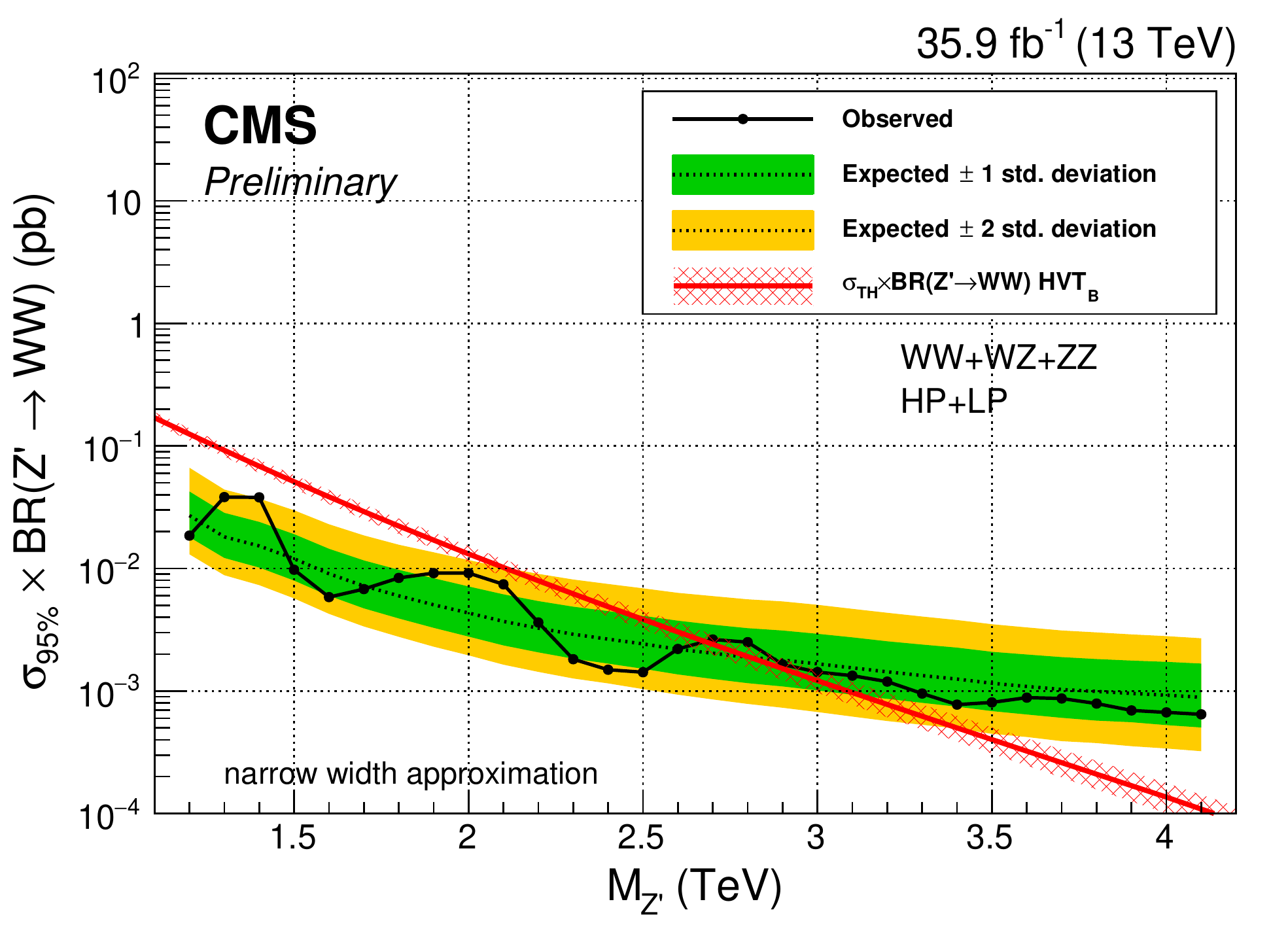}
\includegraphics[height=2in]{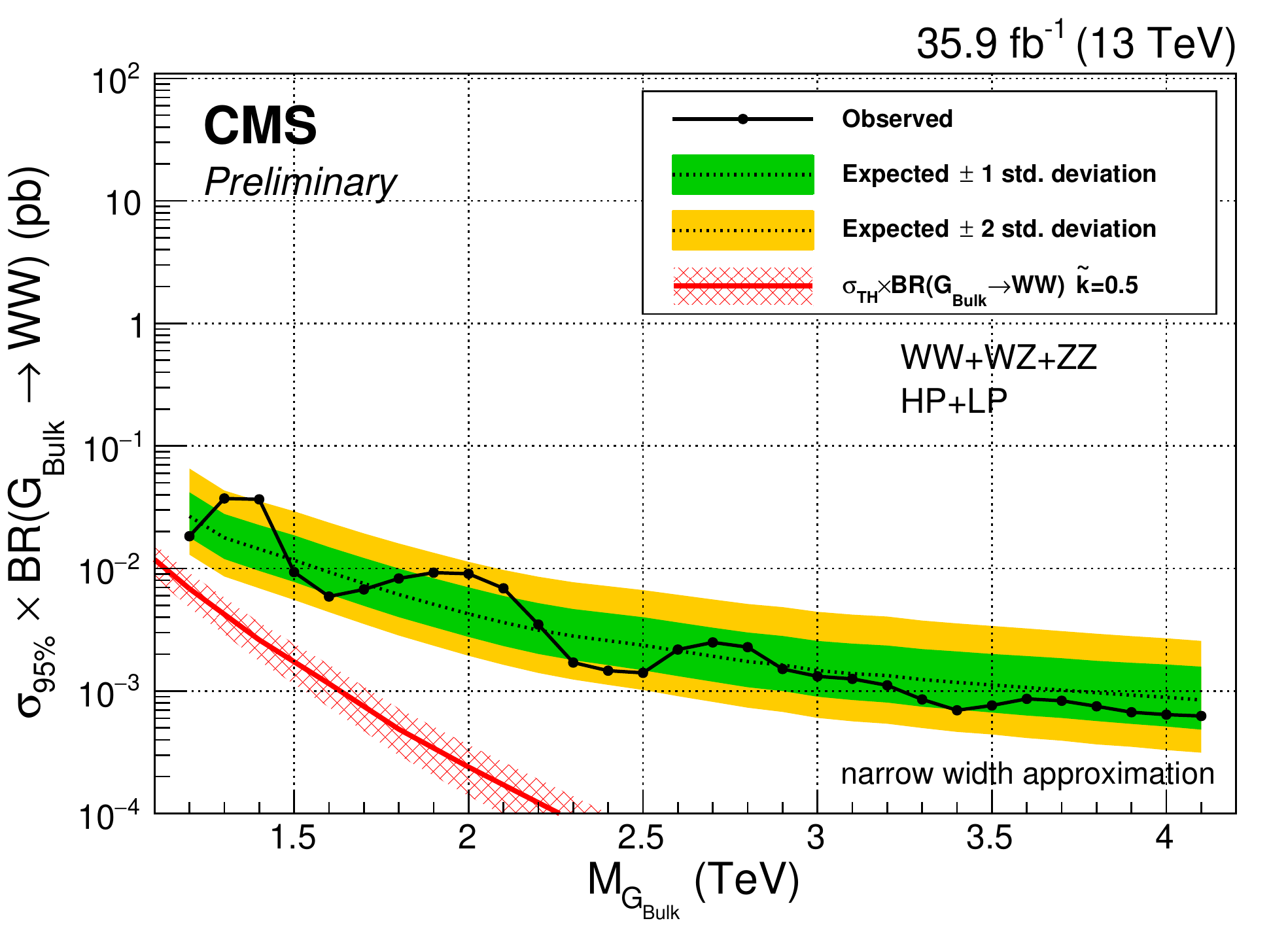}
\caption{The combined observed (black solid) and expected (black dashed) $95\%$ CL upper limits for the spin-1 heavy boson (left) and the spin-2 bulk graviton (right). The solid curves correspond to the cross sections predicted by the HVT and WED models respectively~\cite{VVSearch}.}
\label{fig:VVLimit}
\end{center}
\end{figure}

The background-only hypothesis is tested against the X $\rightarrow$ VV signal in the 6 exclusive categories and is shown in the Z mass window on both jets and the LP $\tau_{21}$ category in Figure~\ref{fig:VVBumpHunt}. The results are interpreted in terms of $95\%$ confidence level (CL) upper limits~\cite{CLs} on the production cross section of the spin-1 heavy boson and the spin-2 graviton and can be seen in Figure~\ref{fig:VVLimit}. 

\section{Search for X $\rightarrow$ VH($\mathrm{q\bar{q}}$$\mathrm{b\bar{b}}$)}

The search for a heavy resonant particle decaying to a vector boson and a Higgs boson is done with the same online and $\Delta\eta$ requirements as the VV search. An additional requirement on the double-b tagger for the H candidate jet is applied. The events are then separated into 8 orthogonal categories, 2 for the b tagging selection for the Higgs jet, 2 categories based on the purity of the $\tau_{21}$ requirement on the V jet, and 2 categories for the mass window on the V jet. 

\begin{figure}[!htb]
\begin{center}
\includegraphics[height=3.in]{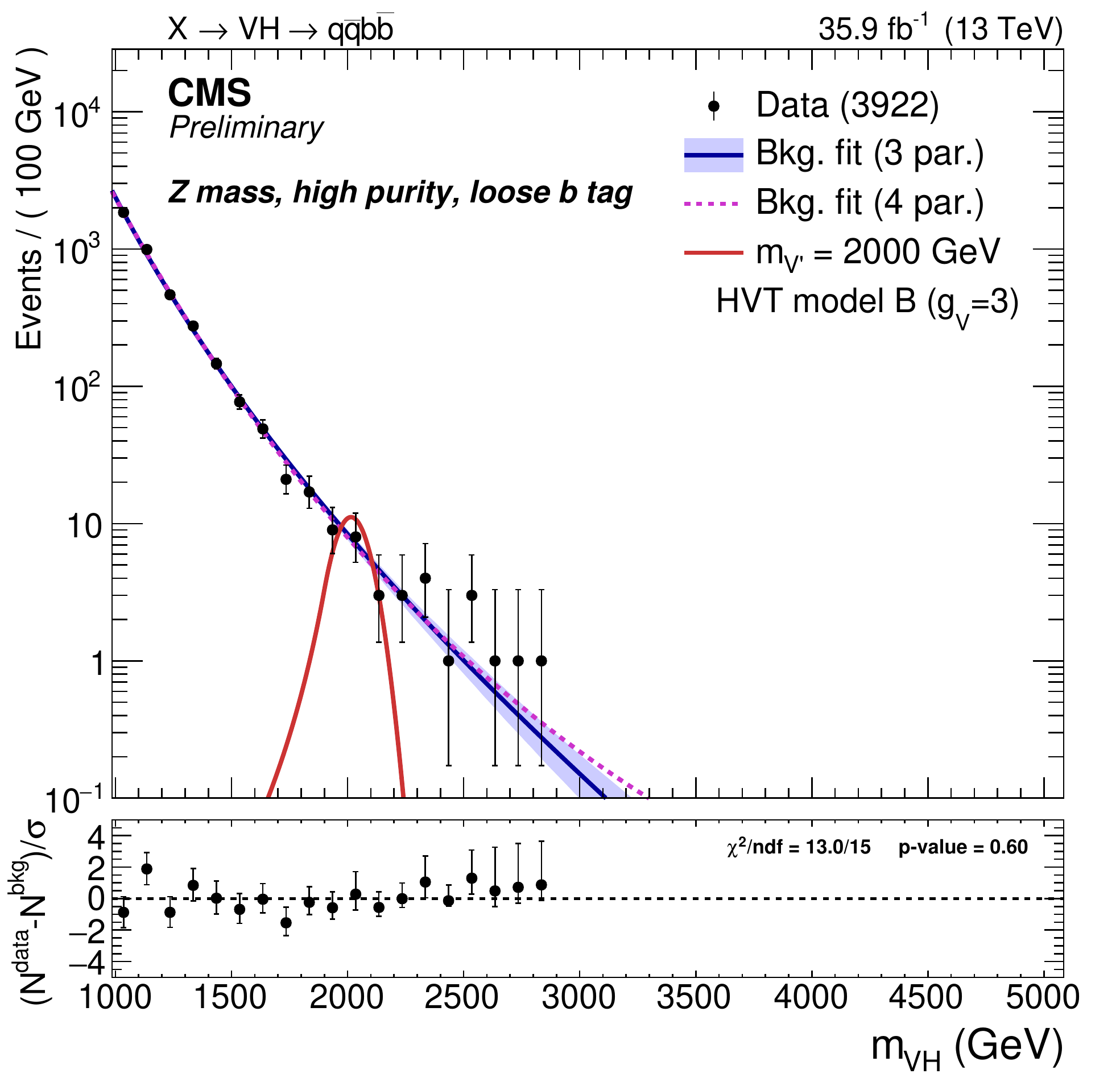}
\caption{
Dijet invariant distribution $m_{VH}$ of the two leading jets in the Z mass region for the high purity category with loose b tagging selections. The preferred background-only fit is shown as a solid blue line with an associated shaded band indicating the uncertainty. An alternative fit is shown as a purple dashed line.  The differences between the data and the predicted background, divided by the data statistical uncertainty ($\sigma$) are shown in the lower panel~\cite{VHSearch}.}
\label{fig:VHBumpHunt}
\end{center}
\end{figure}

\begin{figure}[!htb]
\begin{center}
\includegraphics[height=2in]{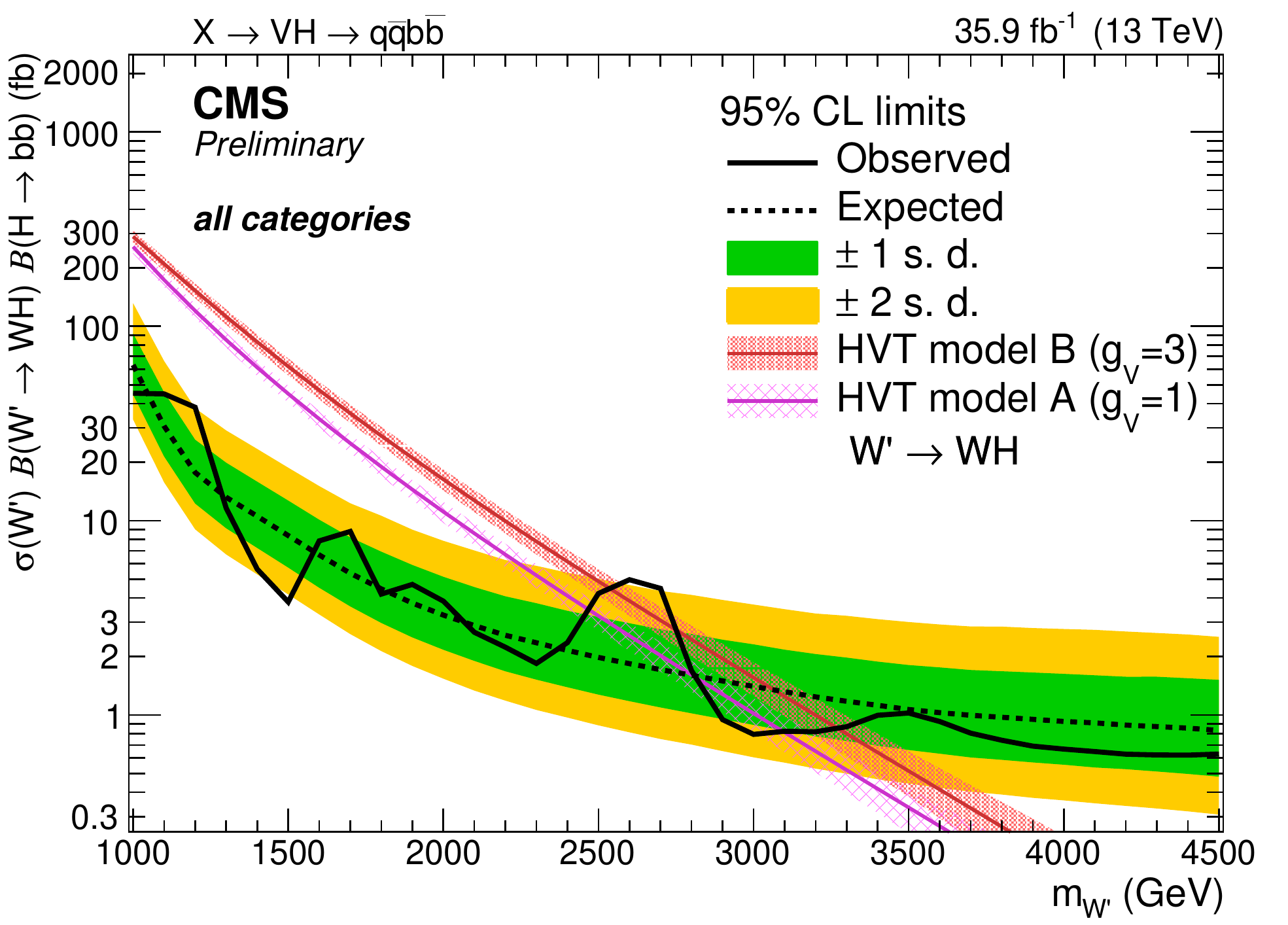}
\includegraphics[height=2in]{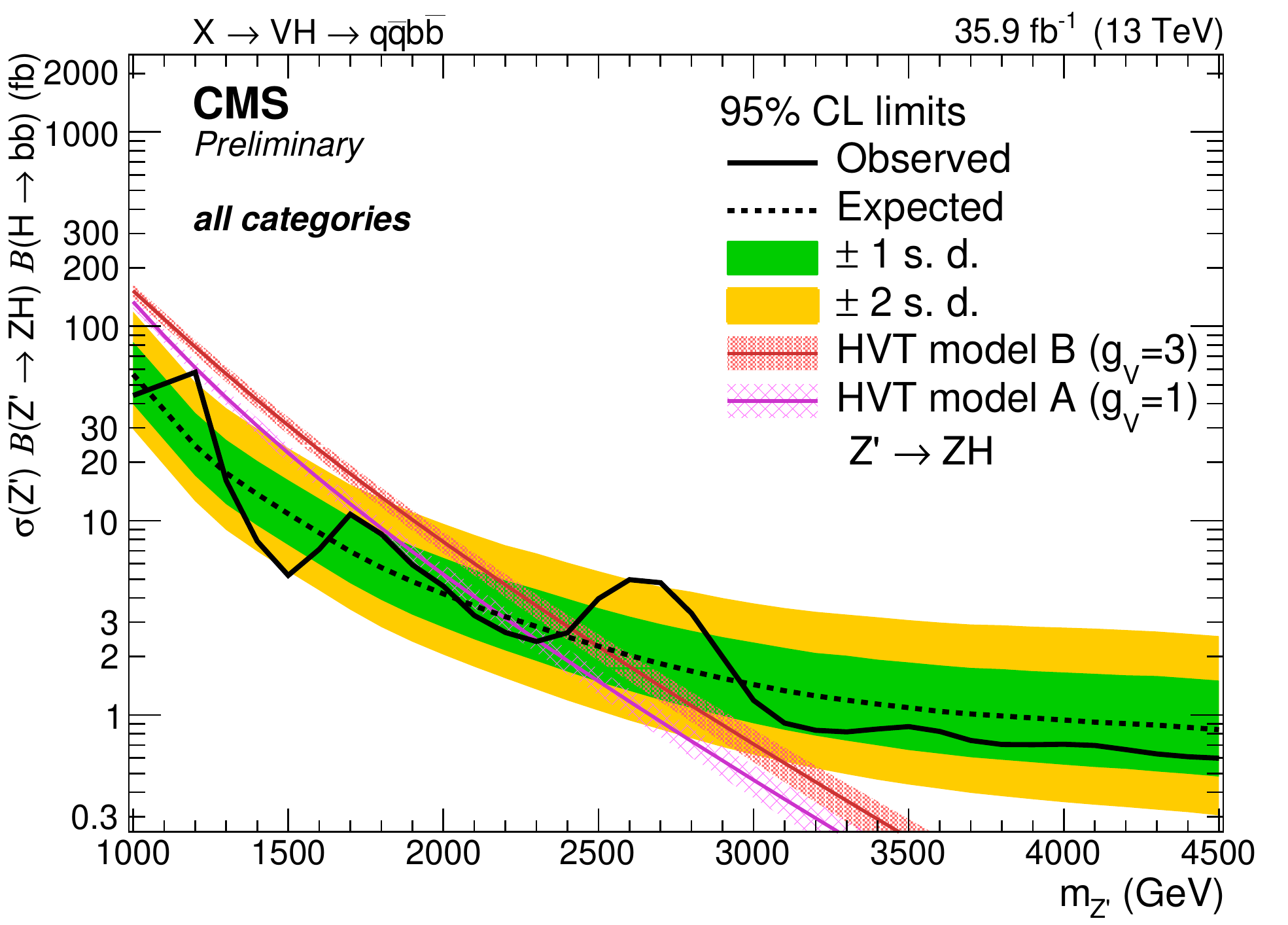}
\caption{The combined observed (black solid) and expected (black dashed) $95\%$ CL upper limits for the $W^{\prime}$ heavy boson (left) and the $Z^{\prime}$ heavy boson (right). The red and purple solid curves correspond to the cross sections predicted by the HVT models~\cite{VHSearch}.}
\label{fig:VHLimit}
\end{center}
\end{figure}

The background is directly estimated from data using the bump hunt methods described for the VV search. The background-only hypothesis is tested against the X $\rightarrow$ VH signal in the 8 exclusive categories and the fit in the Z mass candidate, HP $\tau_{21}$ selection, and loose double-b selection category is shown in Figure~\ref{fig:VHBumpHunt}. The results are interpreted in terms of $95\%$ CL upper limits on the production cross section of the spin-1 heavy boson as seen in Figure~\ref{fig:VHLimit}.

\section{Search for X $\rightarrow$ HH($\mathrm{b\bar{b}}$$\mathrm{b\bar{b}}$)}
The search for a heavy resonant particle decaying to two Higgs bosons is done using the same online and $\Delta\eta$ selection as the VV and VH search. The events are separated into two orthogonal categories that depend on the double-b tagging discriminant of both jets. One category requires each H jet to pass the tight double-b working point (TT) and the other category requires each H jet to pass the loose working point while both don’t pass the tight working point (LL). The TT category achieves good background rejection up to $M_{X}\sim2$ TeV. At higher masses, where the background is small, the LL category helps to recover the signal sensitivity. 

The background is directly estimated from data and is done using two methods depending on whether the triggers are fully efficient, $m_{jj} \geq 1200$ GeV, or not, $m_{jj} < 1200$ GeV. The background estimation in both regions relies on a set of sidebands that are used to predict the total background normalization. Where the triggers are fully efficient the $m_{jj}$ shape is monotonically falling and hence a continuous function is used to model the background while the normalization is constrained from the sidebands.

\begin{figure}[!htb]
\begin{center}
\includegraphics[height=3in]{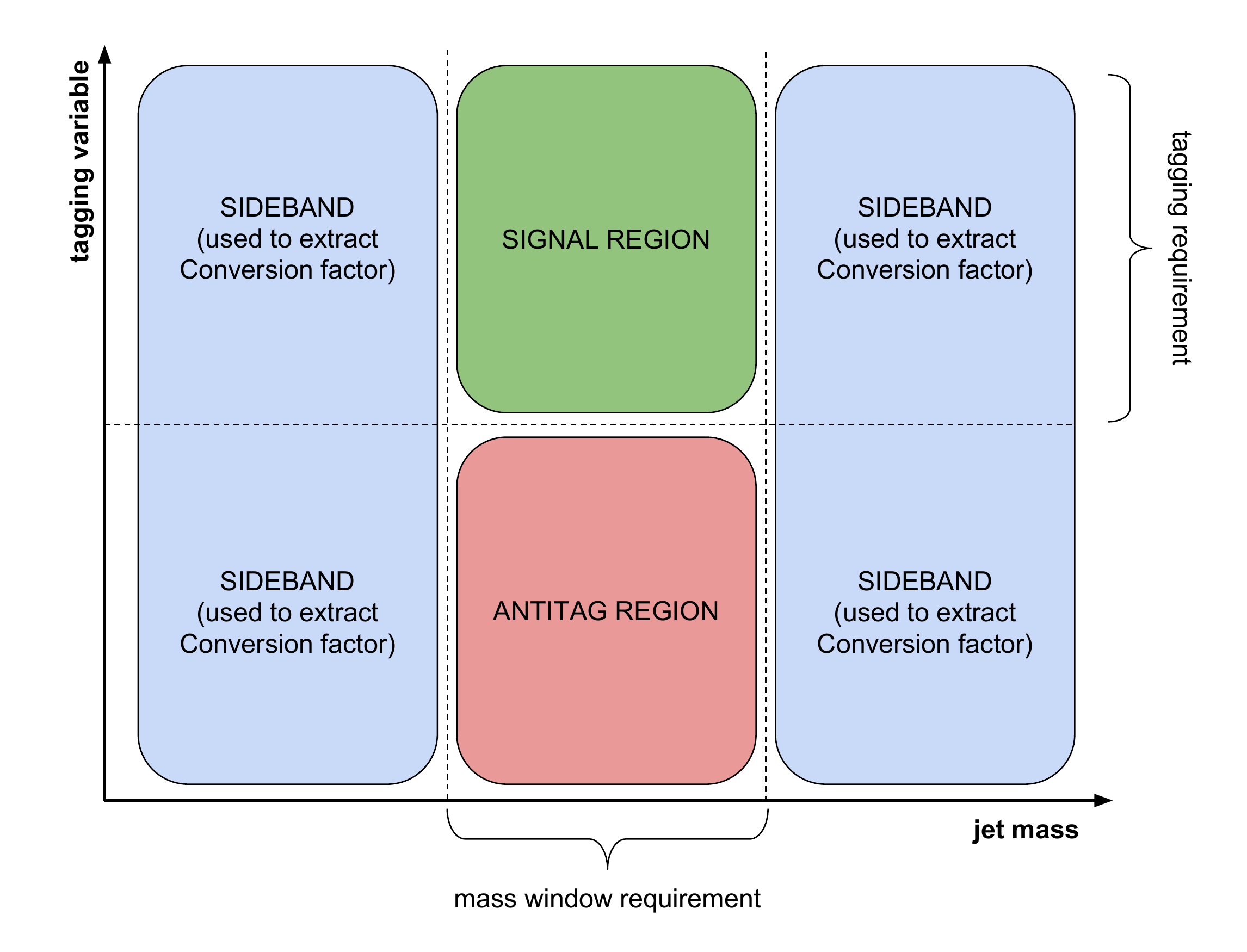}
\caption{Schematic representation of the regions used in the background estimation.}
\label{fig:Alphabet}
\end{center}
\end{figure}

The different sidebands are defined by the soft drop mass, $M_{j1}$, and the double-b discriminant of the leading-$p_{T}$ jet as outlined in Figure~\ref{fig:Alphabet}. The pre-tag region is defined by all events that pass the full selection requirements except the soft drop mass and double-b discriminant requirements on the leading-$p_{T}$ jet. The $M_{j1}$ sideband region consists of events in the pre-tag region, where $M_{j1}$ lies outside the H jet mass window of 105-135 GeV. The signal region is the subset of pre-tag events where $M_{j1}$ is inside the H jet mass window and passing the double-b tagger discriminator cut. The anti-tag region consists of events with $M_{j1}$ within the H jet mass window, but failing the double-b tagger discriminator cut. 

\begin{figure}[!htb]
\begin{center}
\includegraphics[height=2.3in]{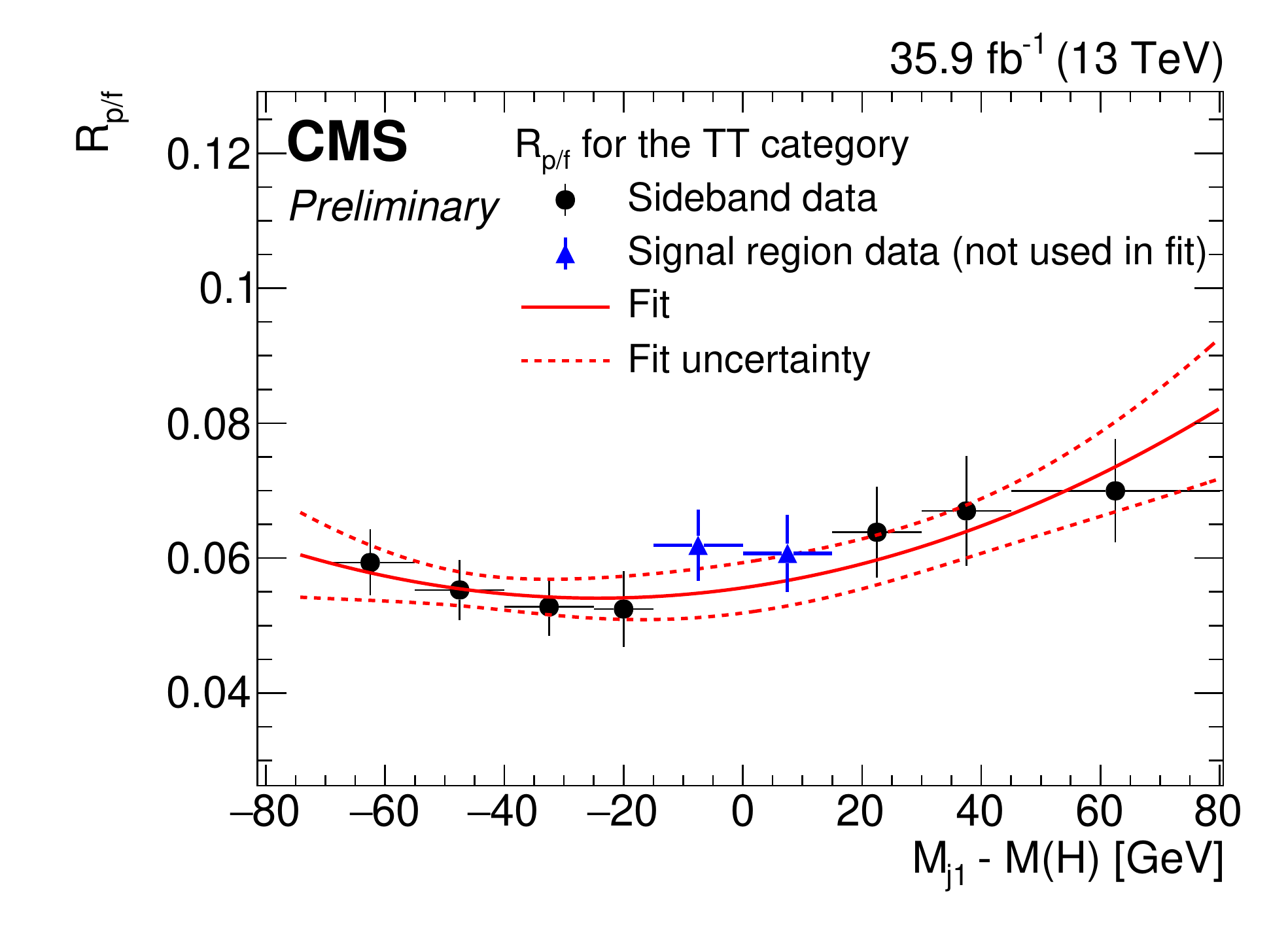}
\includegraphics[height=2.3in]{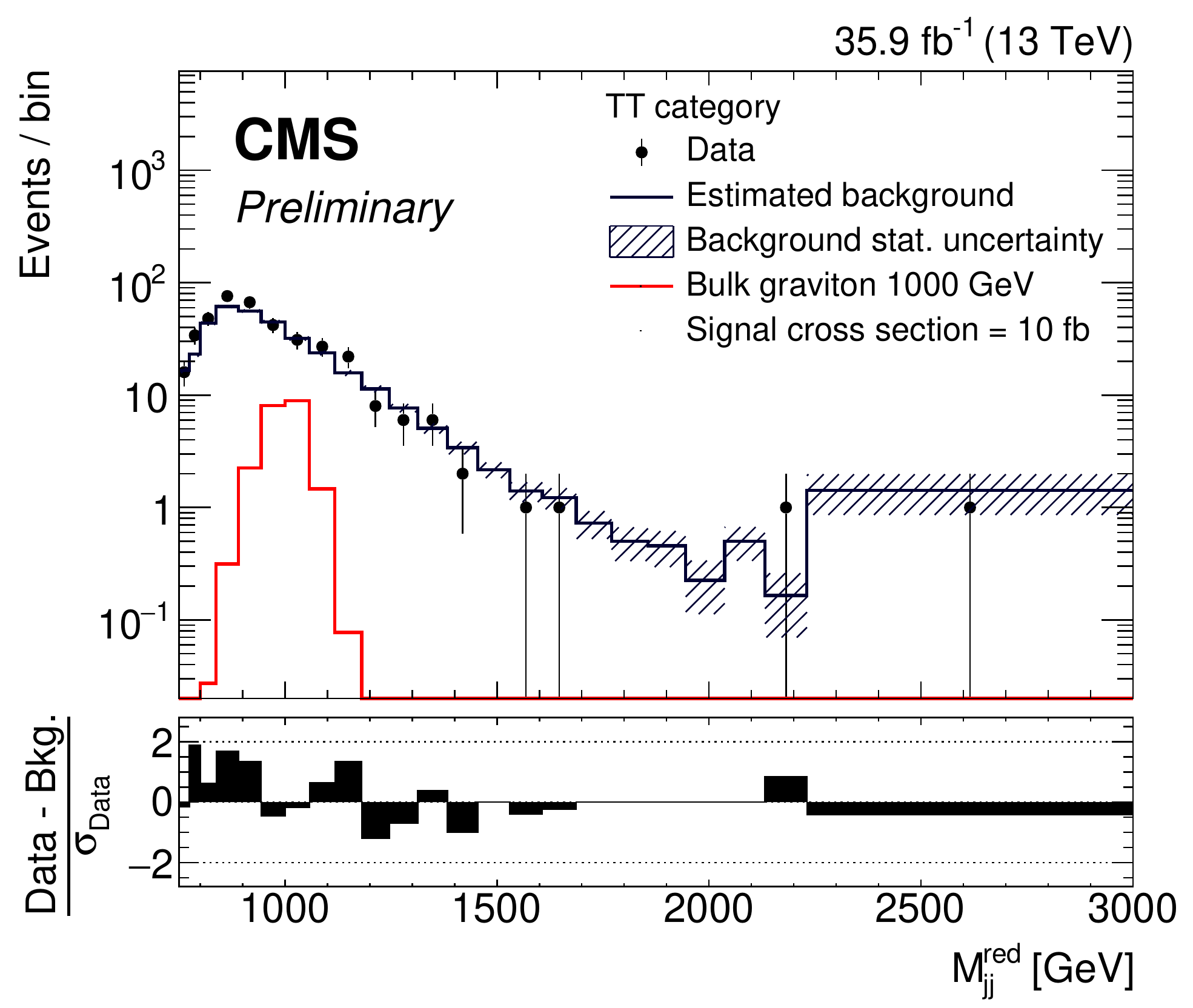}
\caption{
Left: The pass-fail ratio $R_{p/f}$ of the leading-$p_{T}$ jet for the TT signal region categories as a function of $M_{j1} - M_{H}$. Right:  $m_{jj}$ distribution for the TT signal region category. Data points are compared to the estimated background and associated uncertainty. The differences between the data and the predicted background, divided by the data statistical uncertainty (data unc.) are shown in the lower panels~\cite{HHSearch}.}

\label{fig:Alpha}
\end{center}
\end{figure}

In the absence of correlation between $M_{j1}$ and the double-b tagger discriminator values, one could measure in the $M_{j1}$ sideband the ratio of the number of events passing and failing the double-b tagger selection, $R_{p/f}$, and scale the yield in the anti-tag region by this ratio to obtain an estimate of the background normalized in the signal region, \textit{i.e.} the "ABCD" method. To account for the small correlation between the double-b tagger discriminator and $M_{j1}$, the $R_{p/f}$ is measured as a function of $M_{j1}$, thus extending the ABCD method to the "Alphabet" method. The pass-fail ratios in the $M_{j1}$ sidebands are fit with a quadratic function. This way the fit interpolates the pass-fail ratio through the signal region in $M_{j1}$, and every event in the anti-tag region is scaled by the appropriate pass-fail ration given the corresponding $M_{j1}$. The fit to the pass-fail ratios in data as a function of $M_{j1}$ along with the predicted background using this method is shown in Figure~\ref{fig:Alpha}.

\begin{figure}[!htb]
\begin{center}
\includegraphics[height=2.3in]{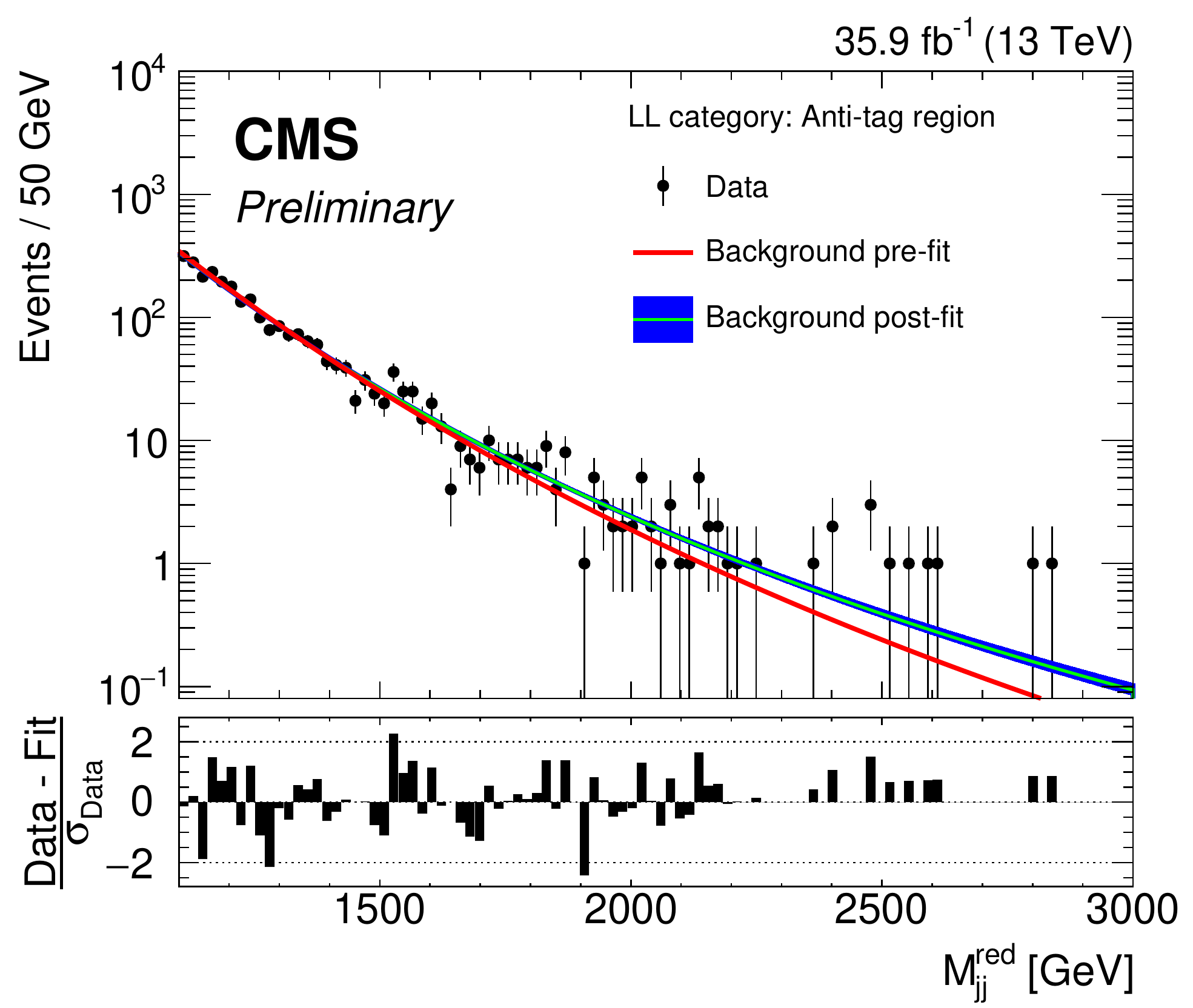}
\includegraphics[height=2.3in]{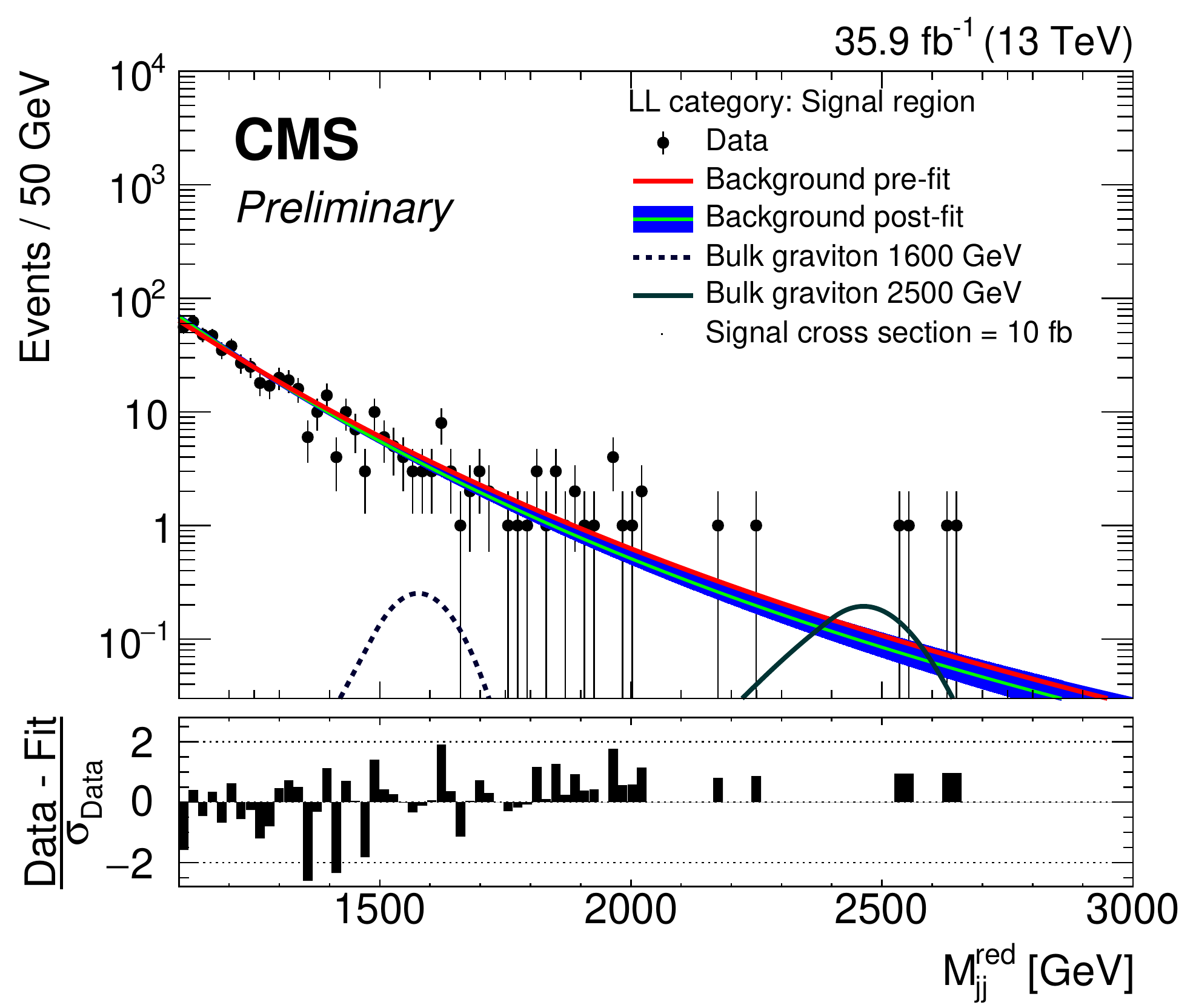}
\caption{
The $m_{jj}$ distributions in the anti-tag (left) and signal region (right) for the LL category. The black markers are the data while the curves show the pre-fit and post-fit background shapes. The differences between the data and the predicted background, divided by the statistical uncertainty in the data ($\sigma_{\mathrm{data}}$) are shown in the lower panels~\cite{HHSearch}.}

\label{fig:AABH}
\end{center}
\end{figure}

\begin{figure}[!htb]
\begin{center}
\includegraphics[height=2in]{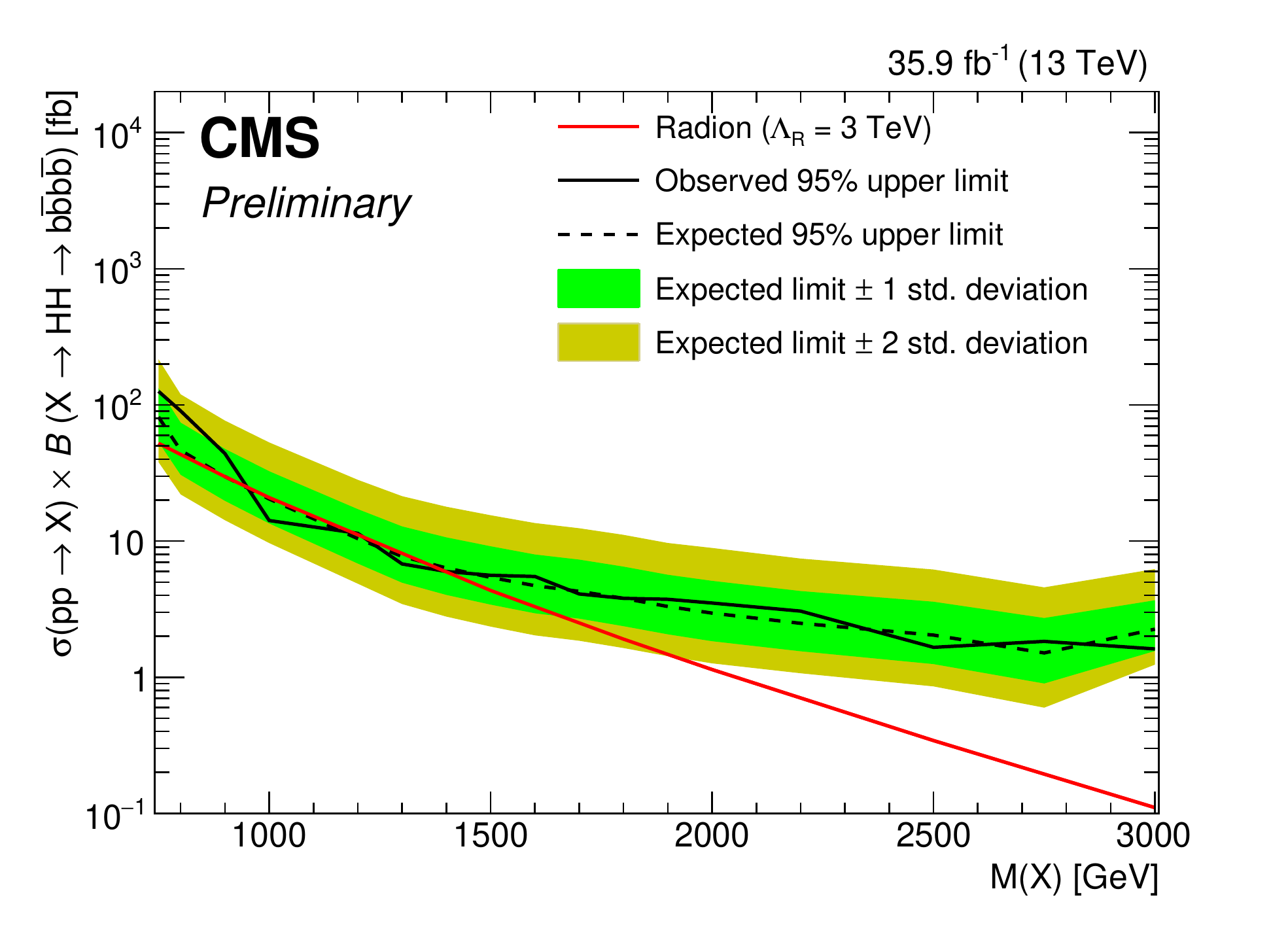}
\includegraphics[height=2in]{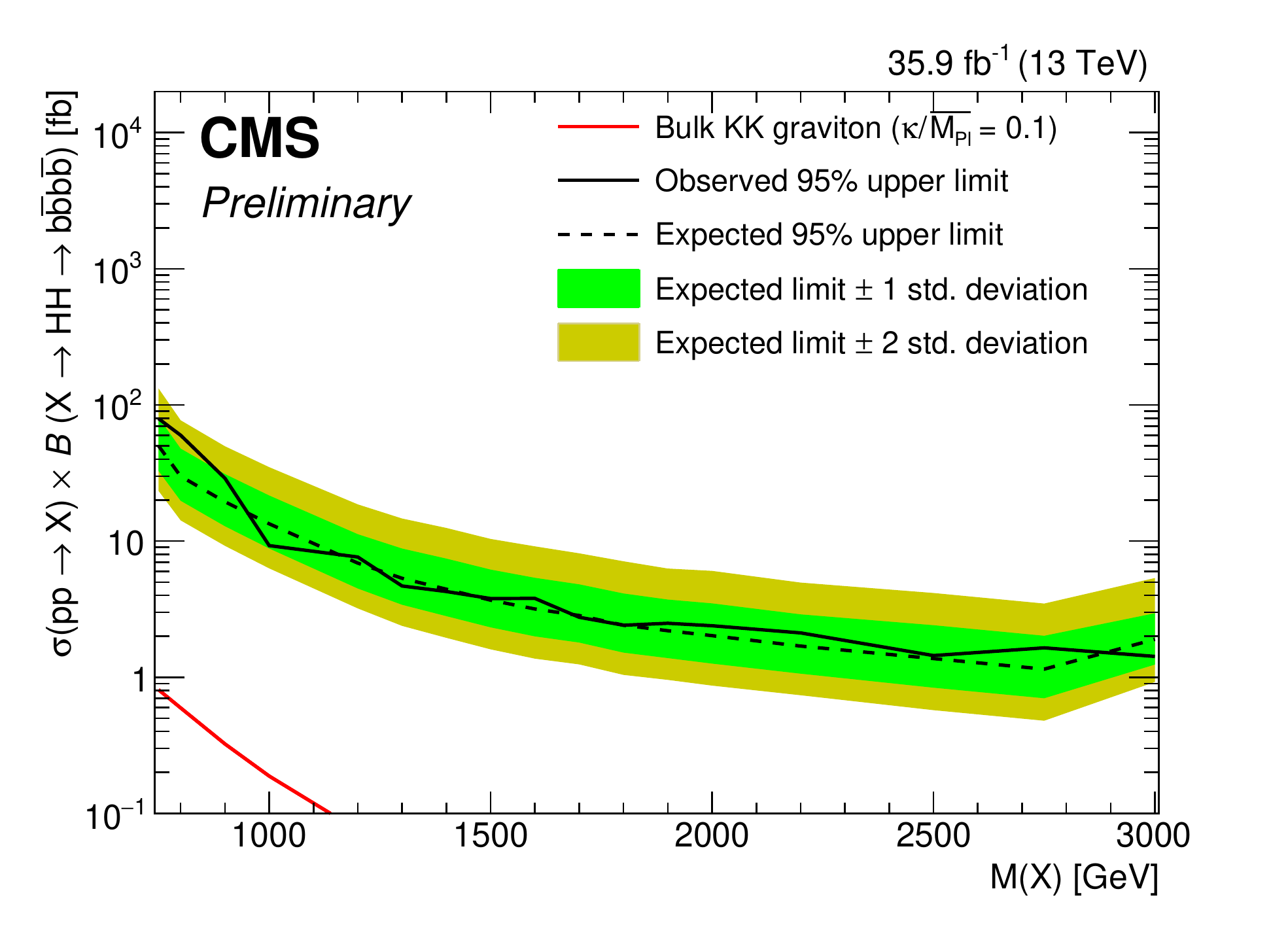}
\caption{The combined limits for the spin-0 radion (left) and the spin-2 bulk graviton (right). The alphabet method is used for mass below 1200 GeV, while the alphabet assisted bump method is used for higher masses. The predicted theoretical cross sections for a narrow radion or a bulk graviton are also shown.~\cite{HHSearch}.}
\label{fig:HHLimit}
\end{center}
\end{figure}

For resonance masses $M_{X} \geq 1200$ GeV, the ratio $R_{p/f}$ obtained from the Alphabet method is used to constrain the number of background events in the signal region while performing the fit. The model used to parameterize the background shape is a 2 parameter levelled exponential function. The results from the fit in the anti-tag and signal region of the LL category are shown in Figure~\ref{fig:AABH}~\cite{HHSearch}. The results are interpreted in terms of $95\%$ CL upper limits on the production cross section of the spin-0 radion and spin-2 graviton as seen in Figure~\ref{fig:HHLimit}.

\section{Summary}

Searches for new massive resonances decaying to two standard model bosons decaying hadronically are reported using data collected with the CMS detector in 2016 at $\sqrt{s}=13$ TeV and corresponding to an integrated luminosity of $35.9\;\mathrm{fb}^{-1}$. The results are interpreted in terms of upper limits on the production cross section of either the spin-0 radion,the spin-1 heavy boson, or the spin-2 graviton. No excess has been found so far, but these searches provide the best upper limits at high resonance masses.

\end{document}